\documentclass[aps,showpacs,twocolumn]{revtex4}
\usepackage{amsfonts}
\usepackage{amsmath}
\usepackage{amssymb}
\usepackage{graphicx}
\usepackage{bm}

\input{tcilatex}

\begin{document}

\title{Thermophoresis of charged colloidal particles}
\author{S\'{e}bastien Fayolle, Thomas Bickel, and Alois W\"{u}rger}
\affiliation{CPMOH, Universit\'{e} Bordeaux 1 \& CNRS, 351 cours de la Lib\'{e}ration,
33405 Talence, France}

\begin{abstract}
Thermally induced particle flow in a charged colloidal suspension is studied
in a fluid-mechanical approach. The force density acting on the charged
boundary layer is derived in detail. From Stokes' equation with no-slip
boundary conditions at the particle surface, we obtain the particle drift
velocity and the thermophoretic transport coefficients. The results are
discussed in view of previous work and available experimental data.
\end{abstract}

\maketitle


\section{Introduction}

A thermally driven flow, or Ludwig-Soret effect, is observed when applying a
temperature gradient to a gasous or liquid phases \cite%
{Gro62,Gid76,Gig77,Wie01,Zhe02,Moh05}. The corresponding mass transport is
relevant for natural and technological processes, such as the global
circulation of sea water~\cite{Cal81} and the phase behavior of eutectic
systems at solidification~\cite{Zhe98}. In recent years, detailed
experimental studies on macromolecular solutions and colloidal suspensions
have revealed unambiguous and often surprising dependencies of the Soret
effect on system parameters such as salinity, surface coating, solute
concentration, and molecular weight~\cite%
{Rau02,Rau05,Wie04,Kit04,Bra02,Zha99,deG03,Iac03,Pia02,Dem04,Duh05,Duh06,Duh06a}%
. Although the analogy to electrophoresis indicates the relevance of surface
forces and suggests a hydrodynamic treatment~\cite{Ruc81,And89}, the
physcial mechanisms that drive thermophoresis in liquids are poorly
understood and differ from those in gaseous phases~\cite{Zhe02,Moh05}.

Applying a generalized force such as an electric field or a thermal gradient
to a complex fluids, results in a flow of heat, charge, particles,... For
sufficiently weak forces, such a non-equilibrium system is described in
terms of linear force-current relations \cite{Gro62}. If the number density $%
n$\ and the temperature $T$ are the relevant variables, the particle current
in a dilute colloidal suspension reads 
\begin{equation}
\mathbf{J}=-D\bm{\nabla}n-nD_{T}\bm{\nabla}T.  \label{2}
\end{equation}%
The first term on the right-hand side corresponds to Fick's law with the
Einstein diffusion coefficient $D$, whereas the second one describes the
thermally induced flow, with the thermal diffusion coefficient $D_{T}$. Eq. (%
\ref{2}) is completed by the expression for the heat current $\mathbf{J}%
_{Q}=-\kappa \bm{\nabla}T-\kappa _{n}\bm{\nabla}n$, with the thermal
conductivity $\kappa $ and the reduced Dufour coefficient $\kappa _{n}$; the
cross-coefficients $\kappa _{n}$ and $D_{T}$ are related by the Onsager
reciprocal rules \cite{Gro62}. The present work is concerned with the
thermophoretic coefficient $D_{T}$ of a charged colloidal suspension.

For a closed system the stationary state is characterized by $\mathbf{J}=0$;
according to Eq. (\ref{2}) a thermal gradient imposes an inhomogeneous
density. Experimentally, $D_{T}$ is determined by applying a temperature
gradient to a uniform suspension ($\bm{\nabla}n=0)$ and by recording the
initial current $\mathbf{J}=-nD_{T}\bm{\nabla}T$, or by measuring the
density modulation $\delta n=-n(D_{T}/D)\delta T$ induced by a temperature
inhomogeneity $\delta T$ in the steady state $\mathbf{J}=0$ \cite{Wie01}.
The latter method gives the Soret coefficient $S_{T}=D_{T}/D$.

Eq.~(\ref{2}) provides a macroscopic description for the particle current~%
\cite{Gro62}. In order to obtain a relation between the kinetic coefficient $%
D_{T}$ and the properties of solute and solvent, we split the particle
current in two terms, 
\begin{equation}
\mathbf{J}=n\mathbf{u}-\mu \bm{\nabla}\Pi \ .  \label{4}
\end{equation}%
The first one accounts for the phoretic velocity $\mathbf{u}$ due to the
interactions at the solute-solvent interface; this is a single-particle
effect, i.e., it is independent of the density $n$ and proportional to the
thermal gradient,%
\begin{equation}
\mathbf{u}=-C\bm{\nabla}T.  \label{4a}
\end{equation}%
The main purpose of this paper is to work out the proportionality factor $C$%
, similar to the coefficients obtained for an electric field or a chemical
gradient \cite{And89}. The second term of $\mathbf{J}$ arises from the
gradient of the osmotic pressure $\Pi $, with the mobility $\mu =1/(6\pi
a\eta )$ depending on the solvent viscosity $\eta $ and on the particle size 
$a$. The stationary state $\mathbf{J}=0$ provides the equilibrium condition
where all forces acting on a given particle cancel. Inserting the
single-particle velocity $\mathbf{u}=-C\bm{\nabla}T$ and the osmotic
pressure of a dilute suspension $\Pi =nk_{B}T$ in (\ref{4}), and comparing
this expression to Eq.~(\ref{2}), we find the Einstein relation $D=\mu
k_{B}T $ and the thermodiffusion coefficient \ 
\begin{equation}
D_{T}=\mu k_{B}+C.  \label{4b}
\end{equation}%
For the Soret coefficient one has 
\begin{equation}
S_{T}=\frac{1}{T}\left( 1+\frac{C}{\mu k_{B}}\right) .  \label{4c}
\end{equation}%
In the absence of particle-solvent interactions one has $C=0$ and $S_{T}=1/T$%
. This simply means that, at constant pressure, the stationary density is
inversely proportional to the non-uniform temperature and that the particles
accumulate in colder regions; this behavior is expected in the absence of
solute-solvent interaction, where the suspended particles may be viewed as
an ideal gas. Yet most colloidal suspensions show a considerably stronger,
positive or negative, Soret effect, i.e., the interaction driven current $-nC%
\bm{\nabla}T$ by far exceeds the ideal-gas term $-n\mu k_{B}\bm{\nabla}T$
and may be directed towards colder or warmer regions. These deviations
express the failure of the ideal-gas picture for the solute and emphasize
the importance of particle-solvent interactions.

The properties of aqueous colloidal suspensions are largely dominated by
charge effects. Besides the surface charge density, the most important
control parameters are the particle radius $a$ and the Debye length\ $%
\lambda $. Recent measurements on suspensions of micelles and polystyrene
nanoparticles \cite{Pia02,Duh06,Duh06a,Vig07,Put07} reported the laws $%
S_{T}\propto a\lambda ^{2}$ or $\propto a^{2}\lambda $, depending on the
experimental conditions and parameters. So far there is no generally
accepted picture for the physical mechanisms at work; theoretical approaches
based on either the free energy of the charged double layer or a
hydrodynamic treatment give diverging results \cite%
{Duh06,Duh06a,Ruc81,Mor99,Par04,Bri03,Fay05,Dho07,Wue07}.

The present work deals with weakly charged particles, in the usual framework
of driven transport in colloidal suspensions \cite{And89}. Sect. 2 gives a
detailed derivation of the force density induced by the thermal gradient in
the vicinity of a charged surface. In Sect.\ 3 we set up the hydrodynamic
description and obtain the fluid and particle velocities; Sect.\ 4 gives the
thermodiffusion coefficient $D_{T}$. In Sect. 5 we compare our results with
previous work and experimental data, and discuss the importance of the
hydrodynamic boundary conditions.

\section{Thermally induced force}

The hydrodynamic treatment given in the following sections relies
essentially on the force $\mathbf{f}dV$ \ exerted by the surface charge of
the particle on a volume element $dV$ of the surrounding fluid. The force
density $\mathbf{f}$ is finite only within a boundary layer of thickness $%
\lambda $. Throughout this paper we suppose that $\lambda $ is much smaller
than the particle radius $a$,%
\begin{equation}
\lambda \ll a.  \label{1e}
\end{equation}%
Thus the hydrodynamic quanitites vary rapidly in the normal direction, and
much more slowly along the interface.

Here we evaluate the electric force density $\mathbf{f}$ that arises from a
thermal gradient.

\subsection{Electrostatics in the boundary layer}

We consider a spherical particle of charge $Q$ and radius $a$. It is
convienent to define the charge density $\sigma =Q/(4\pi a^{2})$. The
surface charge modifies the properties of the fluid in the boundary layer in
several respects. First, it results in an electric field $\mathbf{E}=-%
\mathbf{\nabla }\psi $; the resulting stress is accounted for in terms of
the Maxwell tensor 
\begin{equation*}
\mathcal{T}_{ij}=\varepsilon (E_{i}E_{j}-\frac{1}{2}E^{2}\delta _{ij}).
\end{equation*}%
Second, the electrostatic potential $\psi $ is screened through the
accumulation of mobile counterions in the electrolyte. In mean-field
approximation, the excess densities of (monovalent) positive and negative
ions are given by 
\begin{equation*}
n_{\pm }=n_{0}(e^{\mp e\psi /k_{B}T}-1),
\end{equation*}%
where $n_{0}$ is the salinity. As a result the fluid in the boundary layer
carries a charge density 
\begin{equation*}
\rho =e(n_{+}-n_{-})
\end{equation*}%
and an excess density of mobile ions 
\begin{equation*}
n=n_{+}+n_{-}.\ 
\end{equation*}

Accordingly, the force $\mathbf{f}(\mathbf{r})dV$ acting on a volume element 
$dV$ of the fluid comprises two terms, 
\begin{equation}
\mathbf{f}=\mathbf{\nabla }\cdot \mathcal{T}-\mathbf{\nabla }\left(
nk_{B}T\right) ,  \label{1a}
\end{equation}%
where the divergence of the Maxwell tensor $\mathbf{\nabla }\cdot \mathcal{T}
$ arises from the electric field, and the entropic force $-\mathbf{\nabla }%
\left( nk_{B}T\right) $ from the non-uniform osmotic pressure. The former
term may be rewritten by using the definition of the displacement vector $%
\mathbf{D}=\varepsilon \mathbf{E}$, its relation to the charge density $\rho
=\mathbf{\nabla }\cdot \mathbf{D}$, and the fact that the curl of the
electric field $\mathbf{E}=-\mathbf{\nabla }\psi $ vanishes, $\mathbf{\nabla
\times E}=0$.\ Thus one finds the well-known force density acting on a
charged dielectric body \cite{Str41,Lan83},%
\begin{equation}
\mathbf{\nabla }\cdot \mathcal{T}=\rho \mathbf{E}-\frac{1}{2}E^{2}\mathbf{%
\nabla }\varepsilon ,  \label{1b}
\end{equation}%
where $\rho \mathbf{E}$ describes the action of the electric field, and the
remainder accounts for the dielectrophoretic force due to the spatial
variation of the permittivity \cite{Str41}.\ The additional contribution to
Eq. (\ref{1a}) arises from the osmotic pressure of the mobile ions, i.e.,
from the fact that a charged fluid is a conductor.

\begin{figure}
\includegraphics[width=\columnwidth]{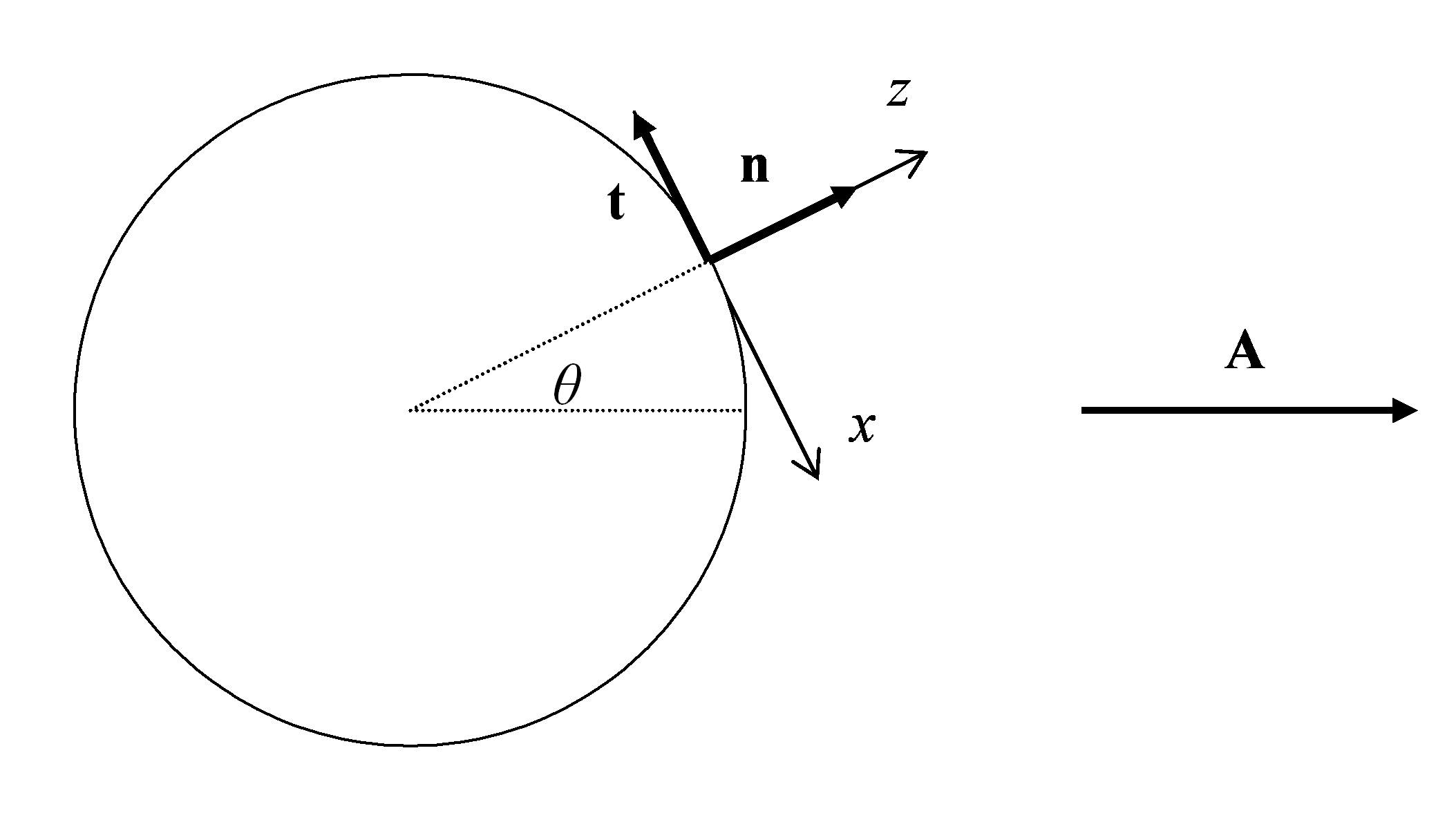}
\caption{Schematic view of a spherical particle of radius $a$, in
a temperature gradient $\mathbf{A}=\bm{\nabla}T$. The surface charge density 
$\protect\sigma $ is screened by a diffuse layer of thickness $\protect%
\lambda \ll a$. At the surface the local coordinates $x,z$, and the normal
and tangent vectors $\mathbf{n}$,$\mathbf{t}$ are indicated. }
\end{figure}

\subsection{Debye-H\"{u}ckel approximation}

The present work is restricted to the\ case of weak surface charges, where
the potential energy of a mobile ion is smaller than the thermal energy, 
\begin{equation}
e\psi \ll k_{B}T.  \label{1d}
\end{equation}
Moreover, (\ref{1e}) implies that the Debye screening length 
\begin{equation*}
\lambda =\sqrt{\frac{\varepsilon k_{B}T}{2n_{0}e^{2}}}
\end{equation*}%
is much smaller than the particle size.

Under the assumption (\ref{1d}) the ion densities $n_{\pm }$ may be expanded
to quadratic order in the small parameter $e\psi /k_{B}T$. Then the charge
density is linear in the potential, $\rho =-\varepsilon \psi /\lambda ^{2}$,
and the electric force in (\ref{1a}) becomes 
\begin{equation*}
\rho \mathbf{E}=\frac{\varepsilon \psi }{\lambda ^{2}}\mathbf{\nabla }\psi =%
\frac{\varepsilon }{2\lambda ^{2}}\mathbf{\nabla }\psi ^{2}.
\end{equation*}%
With the defnition of the Debye length, the leading term of the excess
pressure reads as $nk_{B}T=(\varepsilon \psi ^{2}/2\lambda ^{2})$, and its
gradient may be rewritten as \ 
\begin{equation*}
\mathbf{\nabla }\left( nk_{B}T\right) =\frac{\varepsilon }{2\lambda ^{2}}%
\mathbf{\nabla }\psi ^{2}-\frac{\varepsilon \psi ^{2}}{2\lambda ^{2}}\frac{%
\mathbf{\nabla }T}{T}.
\end{equation*}%
Inserting these terms in (\ref{1a}) we get 
\begin{equation}
\mathbf{f}=-\frac{1}{2}E^{2}\mathbf{\nabla }\varepsilon +\frac{\varepsilon
\psi ^{2}}{2\lambda ^{2}}\frac{\mathbf{\nabla }T}{T}.  \label{7a}
\end{equation}%
So far we have not used the precise form of the electrostatic potential. The
force $\mathbf{f}$ solely depends on the gradients of the permittivity $%
\varepsilon $ and the non-uniform temperature $T(\mathbf{r})$.

Since this work is confined to the case of a thin boundary layer, $\lambda
\ll a$, we use the screened electrostatic potential of a flat surface 
\begin{equation}
\psi =\psi _{0}e^{-z/\lambda },\ \ \ \ \ \ \ \psi _{0}=\frac{\lambda \sigma 
}{\varepsilon }.  \label{8}
\end{equation}%
Taking the derivative with respect to the normal coordinate gives the
electric field $E=\psi /\lambda $.\ The charge density of the fluid reads $%
\rho =-(\sigma /\lambda )e^{-z/\lambda }$, one readily verifies $\int dz\rho
=-\sigma $, i.e., the overall charge of the double layer is zero.

Comparison with the surface potential of a spherical particle $Q/[4\pi
\varepsilon (1+a/\lambda )a]=\psi _{0}/(1+\lambda /a)$ reveals that the
finite curvature would result in corrections of the order $\lambda /a$.\
Since we rely on Eq. (\ref{1e}) throughout this paper, these corrections are
of no significance.

\subsection{Uniform temperature}

We briefly address the force balance in a uniform system where the
permittivity and the temperature are constants $\mathbf{\nabla }T=0$.
Clearly, Eq. (\ref{7a})\ states that the total force vanishes, $\mathbf{f}=0$%
; yet each of the terms in Eq. (\ref{1a}) contains a finite contributions
that is independent of $\mathbf{\nabla }T$.\ From the potential (\ref{8})\
one finds the force on the charged fluid 
\begin{equation*}
\rho E_{z}=-\rho \partial _{z}\psi
\end{equation*}%
and the gradient of the entropic pressure 
\begin{equation*}
k_{B}T\partial _{z}n=(\varepsilon /\lambda ^{2})\psi \partial _{z}\psi .
\end{equation*}%
With the above expression for the charge density these terms cancel each
other. This just means that for $\mathbf{\nabla }T=0$\ there is no net force
and that the fluid is immobile.

\subsection{Thermal force}

Now we consider the effect of a small but finite temperature gradient. Since
the permittivity gradient in Eq. (\ref{7a}) arises from the nonuniform
temperature, $\mathbf{\nabla }\varepsilon =(d\varepsilon /dT)\mathbf{\nabla }%
T$, both contributions to the thermal force are already proportional to $%
\mathbf{\nabla }T$. In view of the linear current-force relation (\ref{2}),
terms of higher order in $\mathbf{\nabla }T$ are irrelevant. Thus we
calculate the coefficients in (\ref{7a}) from the unperturbed potential (\ref%
{8}). In particular, this leads to the electric field $E=\psi /\lambda $;
with the logarithmic derivative of the permittivity 
\begin{equation*}
\tau =-\frac{d\ln \varepsilon }{d\ln T}
\end{equation*}%
we obtain the final form for the force field 
\begin{equation}
\mathbf{f}=\frac{\varepsilon \psi ^{2}}{2\lambda ^{2}}\left( 1+\tau \right) 
\frac{\mathbf{\nabla }T}{T}.  \label{7}
\end{equation}

\subsection{Temperature gradient}

Due to the different heat conductivities of particle and solvent, the
temperature field close to the particle surface differs from the externally
applied uniform modulation \cite{Gid95}. In the introduction, e.g., in Eq. (%
\ref{4a}), $\mathbf{\nabla }T$ refers to the externally applied thermal
gradient, i.e., to its value far from the particle, which we denote 
\begin{equation*}
\mathbf{A}=\left. \bm{\nabla}T\right\vert _{\infty }
\end{equation*}%
in the remainder of this paper. On the other hand, Eq. (\ref{7}) involves
the gradient close to the particle surface. Since heat propagation is much
faster than particle migration, the temperature field may be taken as
stationary. The heat conduction equation for a spherical particle is readily
solved, and the tangential component of the thermal gradient at the particle
surface reads~\cite{Lan87} 
\begin{equation}
\partial _{x}T=-\xi (\mathbf{t}\cdot \mathbf{A})=\xi A\sin \theta .
\label{11}
\end{equation}%
As shown in Fig. 1, $\theta $ denotes the angle between the surface normal $%
\mathbf{n}$ and the applied gradient $\mathbf{A}$. According to the usual
definition of polar coordinates, the tangent vector $\mathbf{t}$ coincedes
with the negative $x$-axis. The parameter $\xi =3\kappa _{S}/(2\kappa
_{S}+\kappa _{P})$ is determined by the ratio of the heat conductivities of
solvent and particle. As to the the normal component, one finds 
\begin{equation*}
\partial _{z}T=\xi _{n}(\mathbf{n}\cdot \mathbf{A})=\xi _{n}A\cos \theta
\end{equation*}%
with a modified prefactor $\xi _{n}=3\kappa _{P}/(2\kappa _{S}+\kappa _{P})$.

The unperturbed temperature gradient reads in local coordinates $A_{x}=A\sin
\theta $ and $A_{z}=A\cos \theta $. Thus the changes at the surface of a
colloidal particle are expressed by the factors $\xi $ and $\xi _{n}$. For
the case where the heat conductivities of solute and solvent are identical, $%
\kappa _{S}=\kappa _{P}$, we have $\xi =1=\xi _{n}$, i.e., the thermal
gradient is constant everywhere, $\bm{\nabla}T=\mathbf{A}$.

\section{Hydrodynamics}

The particle velocity $\mathbf{u}$ has to be derived from a fluid-mechanical
treatment~\cite{And89}. The thermally driven motion of micron or nanometer
sized particles in a viscous liquid involves small Reynolds numbers, i.e.,
inertia effects are negligible. Then the stationary velocity is given by
Stokes' equation~\cite{Lan87} 
\begin{equation}
\eta \nabla ^{2}\mathbf{v}=\bm{\nabla}P-\mathbf{f},  \label{1}
\end{equation}%
where $\eta $ is the solvent viscosity, $P$ the hydrostatic pressure, and $%
\mathbf{f}$ the force density exerted by the particle on the fluid. An
incompressible fluid satisfies $\bm{\nabla}\cdot \mathbf{v}=0$, and in
general stick boundary conditions are supposed to apply at the particle
surface, 
\begin{equation}
\mathbf{v|}_{r=a}=\mathbf{u}.  \label{1c}
\end{equation}

The characteristic length scales of the normal and parallel derivatives in
Eq.~(\ref{1}) are given by $\lambda $ and $a$. The condition of a thin
boundary layer, as expressed in Eq. (\ref{1e}), thus implies that the forces
vary rapidly in the normal direction, and much more slowly along the
interface. The resulting separation of length scales permits us to calculate
the particle velocity in two steps. First, resorting to a 1D approximation
that is valid at distances much shorter than the particle size, we derive
the boundary velocity induced by the thermal force. In a second step we
match this solution with that of the force-free Stokes equation at distances
well beyond $\lambda $, and thus obtain the fluid velocity field.

In the laboratory frame the particle moves at speed $\mathbf{u}$ and the
fluid velocity vanishes at infinity. For the sake of computational
simplicity, we transform to the reference frame in which the particle is at
rest. Indicating the corresponding velocities by a hat, we have $\mathbf{%
\hat{u}}=0$ and 
\begin{equation*}
\mathbf{\hat{v}}(\mathbf{r})=\mathbf{v(\mathbf{r}})-\mathbf{u}.
\end{equation*}%
For the fluid motion in the boundary layer, local coordinates $x$ and $z$
turn out to be most convenient, whereas the velocity field at larger
distances is best described in terms of polar coordinates $r$ and $\theta $\
with the origin at the particle center. See Fig.\ 1.

\subsection{Boundary layer}

We rewrite (\ref{1}) in terms of the normal and parallel coordinates $z$ and 
$x$, 
\begin{equation*}
\eta \mathbf{\nabla }^{2}\hat{v}_{i}=\partial _{i}P-f_{i},
\end{equation*}%
with $\partial _{x}=\partial /\partial x$, etc., and where $\mathbf{\hat{v}}$
is the relative fluid velocity with respect to the particle surface. The
normal component vanishes close to the interface, $\hat{v}_{z}=0$, which
implies $\partial _{z}P-f_{z}=0$ \cite{And89}. Since the force $\mathbf{f}$
is finite within the boundary layer only, the hydrostatic pressure is
constant at larger distances. Integrating $\partial _{z}P=f_{z}$ we have 
\begin{equation}
P=P_{0}-\int_{z}^{\Lambda }dz^{\prime }f_{z}(z^{\prime }),  \label{1f}
\end{equation}%
where $P_{0}$ is a constant. The upper bound of the integral is much larger
than the thickness of the boundary layer but much smaller than the particle
size, and thus satisfies 
\begin{equation*}
\lambda \ll \Lambda \ll a.
\end{equation*}

Regarding the tangential velocity $\hat{v}_{x}$, its derivative along the
surface is much smaller than the normal component, resulting in the
inequality $\partial _{x}^{2}\hat{v}_{x}\ll \partial _{z}^{2}\hat{v}_{x}$.
Discarding $\partial _{x}^{2}\hat{v}_{x}$ accordingly, the equation for the
tangential velocity component becomes 
\begin{equation}
\eta \partial _{z}^{2}\hat{v}_{x}=\partial _{x}P-f_{x}.  \label{5}
\end{equation}%
The derivative on the right-hand side gives the lateral force per unit
volume exerted on the fluid.\ Integrating this relation once gives the shear
stress $\sigma _{xz}=\eta \partial _{z}\hat{v}_{x}$. This quantity does not
have a rigorous reference value, i.e., it takes finite values both at the
particle surface and beyond the boundary layer. Yet in the boundary layer
approximation, i.e., by assuming an infinite flat surface, $\sigma _{xz}$ is
zero well beyond the boundary layer. Taking $\sigma _{xz}(\Lambda )=0$ as
reference value, the shear stress is given by its variation from $\Lambda $
to a distance $z$ from the surface, 
\begin{equation}
\sigma _{xz}(z)=\int_{z}^{\Lambda }dz^{\prime }(f_{x}-\partial _{x}P).
\label{5a}
\end{equation}%
Integrating once more gives the velocity of the fluid with respect to the
particle, 
\begin{equation}
\hat{v}_{x}(z)=\frac{1}{\eta }\int_{0}^{z}dz^{\prime }\sigma _{xz}(z^{\prime
}).  \label{5c}
\end{equation}%
Here we have used stick boundary conditions, i.e., $\hat{v}_{x}(z)$ is zero
at $z=0$.

\subsection{Boundary velocity}

Replacing the upper bound of the integral with $z\rightarrow \Lambda $, the
quantity $v_{x}(\Lambda )$ gives the relative velocity of the fluid beyond
the boundary layer with respect to the particle surface. Inserting the
non-uniform pressure $P$ one finds \ 
\begin{equation}
v_{B}=\frac{1}{\eta }\int_{0}^{\Lambda }dz\int_{z}^{\Lambda }dz^{\prime
}\left( f_{x}+\frac{\partial }{\partial x}\int_{z^{\prime }}^{\Lambda
}dz^{\prime \prime }f_{z}(z^{\prime \prime })\right) .  \label{9}
\end{equation}%
On a mesoscopic level the relevant length scale is given by the particle
size\ $a$; because of $\Lambda \ll a$, one may consider the limit $\Lambda
/a\rightarrow 0$ and take $v_{B}$ as the fluid velocity at the interface.

This velocity depends on both components of the force density $f_{x}$ and $%
f_{z}$. We show that for the electric force studied here, the latter
contribution is negligible, i.e., the tangential derivative $\partial _{x}P$
of the pressure is small as compared to $f_{x}$. Indeed, if the thermal
conductivities of solvent and particle are not very different, the normal
and parallel temperature gradients $\partial _{x}T$ and $\partial _{z}T$
components are of the same order of magnitude, and so are the force
components $f_{x}$ and $f_{z}$. Since the force is finite within the
boundary layer only, the second term in parentheses in (\ref{9}) is
approximately $\partial _{x}(\lambda f_{z})$; this has to be compared with $%
f_{x}$. From (\ref{7}) it is clear that $\partial _{x}(\lambda f_{z})$
comprises terms proportional either to the square of the thermal gradient $%
\propto (\partial _{x}T)(\partial _{z}T)$ or to the second derivative $%
\propto \partial _{x}\partial _{z}T$. (The curvature of the temperature
field vanishes in the bulk fluid, but is finite in the boundary layer.) The
quadratic terms arise from the factors $\lambda ,\varepsilon ,\psi ,T$
present in $\lambda f_{z}$; they are not significant in view of the linear
current-force relation (\ref{2}). As to the second derivative, the above
discussion of the thermal gradient implies that $\partial _{x}\partial _{z}T 
$ varies on the scale of the particle size, $\partial _{x}\partial _{z}T\sim
(1/a)\partial _{z}T$. Thus we find that the second term in parenthesis in (%
\ref{9}) is at most of the order, 
\begin{equation*}
\partial _{x}(\lambda f_{z})\sim (\lambda /a)f_{z}\ll f_{x}.
\end{equation*}%
As a consequence of this \textquotedblleft boundary layer
approximation\textquotedblright\ \cite{And89}, we discard the integral term
and have 
\begin{equation*}
v_{B}=\frac{1}{\eta }\int_{0}^{\Lambda }dz\int_{z}^{\Lambda }dz^{\prime
}f_{x}.
\end{equation*}

Inserting the tangential component $f_{x}$ in (\ref{9}) and noting that only
the electrostatic potential $\psi $ depends on the integration variable, we
obtain a double integral of $\psi ^{2}$. With the above expression for $\psi 
$ this integral is readily performed; putting $e^{-\Lambda /\lambda
}\rightarrow 0$ one finds 
\begin{equation*}
\int_{0}^{\Lambda }dz\int_{z}^{\Lambda }dz^{\prime }\psi ^{2}=\frac{\lambda
^{2}\psi _{0}^{2}}{4},
\end{equation*}%
and thus the boundary velocity 
\begin{equation*}
v_{B}=\frac{\varepsilon \psi _{0}^{2}}{8\eta T}\left( 1+\tau \right)
\partial _{x}T.
\end{equation*}%
The parallel component of the temperature gradient depends on the
orientiation of the surface with respect to $\mathbf{A}$. Separating the
resulting sine function, we have 
\begin{equation*}
v_{B}=v_{0}\sin \theta ,
\end{equation*}%
with the constant 
\begin{equation}
v_{0}=\frac{\varepsilon \psi _{0}^{2}}{8\eta }\left( 1+\tau \right) \frac{%
\xi A}{T}.  \label{10a}
\end{equation}%
For later use we give the vector quantity in the basis related to polar
coordinates, 
\begin{equation}
\mathbf{v}_{B}=-v_{0}\sin \theta \ \mathbf{t},  \label{10}
\end{equation}%
where the minus arises since the tangent vector $\mathbf{t}$ points along
the negative $x$-axis.

\subsection{Three-dimensional flow}

The electric force and the boundary velocity have been evaluated in terms of
a one-dimensional approximation to Stokes' equation that ceases at distances
beyond $\Lambda $. In this range one has to deal with the 3D Stokes
equation, albeit with modified boundary conditions. In a mesoscopic
description, we may put $\Lambda /a\rightarrow 0$ and consider $v_{B}$ as
the fluid velocity at the interface. Thus Eq.~(\ref{1}) reduces to the
force-free Stokes equation 
\begin{equation}
\eta \mathbf{\nabla }^{2}\mathbf{v}=\bm{\nabla}P.  \label{12}
\end{equation}

Treating the fluid as incompressible imposes continuity of the normal
component of the velocity, 
\begin{equation}
\mathbf{n}\cdot (\mathbf{v}-\mathbf{u})\mathbf{|}_{a+\Lambda }=0.  \label{15}
\end{equation}%
A second condition is obtained by noting that there is no net external force
acting on the system consisting of the particle and the charged fluid. Thus
the integrated normal force outside the boundary layer vanishes~\cite{And89}%
, 
\begin{equation}
{\displaystyle\oint }_{a+\Lambda }dS\bm{\sigma}\cdot \mathbf{n}=0\ .
\label{16}
\end{equation}%
where the stress tensor 
\begin{equation*}
\bm{\sigma}=\bm{\sigma}^{\prime }-P
\end{equation*}%
comprises the dissipative term or viscous force density $\sigma
_{ij}^{\prime }=\eta \left( \partial _{i}v_{j}+\partial _{j}v_{i}\right) $,
and the hydrostatic pressure $P$~\cite{Lan87}. The third condition involves
the velocity~(\ref{10}), which accounts for the force acting on the double
layer, 
\begin{equation}
\mathbf{t}\cdot (\mathbf{v}-\mathbf{u})\mathbf{|}_{a+\Lambda }=\mathbf{t}%
\cdot \mathbf{v}_{B}.  \label{17}
\end{equation}

We transform to the reference frame in which the particle is at rest, with $%
\mathbf{\hat{u}}=0$ and $\mathbf{\hat{v}}(\mathbf{r})=\mathbf{v(\mathbf{r}})-%
\mathbf{u}$. The solution of Stokes' equation at small Reynolds numbers in
spherical coordinates $\mathbf{\hat{v}}=\hat{v}_{r}\mathbf{n}+\hat{v}%
_{\theta }\mathbf{t}$ reads~\cite{Lan87} 
\begin{subequations}
\begin{align}
\hat{v}_{r}& =-u\cos \theta \left( 1-2\alpha \frac{a}{r}+2\beta \frac{a^{3}}{%
r^{3}}\right) \ , \\
\hat{v}_{\theta }& =u\sin \theta \left( 1-\alpha \frac{a}{r}-\beta \frac{%
a^{3}}{r^{3}}\right) \ ,  \label{6}
\end{align}%
where $\theta $ is the polar angle with respect to the $x$ axis and the
radial and tangential unit vectors $\mathbf{n}=\mathbf{r}/r$ and $\mathbf{t}%
=\partial \mathbf{n}/\partial \theta $. This flow field is related to a
non-uniform hydrostatic pressure 
\end{subequations}
\begin{equation*}
P(\mathbf{r})=P_{0}+\alpha \frac{2\eta ua}{r^{2}}\cos \theta \,.
\end{equation*}%
The parameters $u,\alpha ,\beta $ are determined from the solulution of
Stokes' with the boundary conditions~(\ref{15})--(\ref{17}). \ 

The first two of these conditions involve the fluid velocity and stress
only. In the particle-fixed frame the normal velocity vanishes, $\hat{v}%
_{r}|_{r=a}=0$, resulting in $1-2\alpha +2\beta =0$. The total stress at the
interface can be written as $\bm{\sigma}\cdot \mathbf{n}=\mathbf{n}\sigma
_{rr}+\mathbf{t}\mathbb{\sigma }_{r\theta }$, with the entries of the
dissipative part in shperical coordinates \cite{Lan87} 
\begin{equation}
\sigma _{rr}^{\prime }=2\eta \frac{\partial \hat{v}_{r}}{\partial r},\ \ \
\sigma _{r\theta }^{\prime }=\eta \left( \frac{\partial \hat{v}_{\theta }}{%
\partial r}-\frac{\hat{v}_{\theta }}{r}\right) \ .  \label{5b}
\end{equation}%
Integrating~(\ref{16}) over a sphere just outside the boundary layer gives
the relation $1-5\alpha +2\beta =0$. One readily obtains the amplitudes of
the velocity field varying with distance as $1/r$ and $1/r^{3}$,
respectively, 
\begin{equation}
\alpha =0,\ \ \ \ \ \beta =-\frac{1}{2}\ .  \label{19}
\end{equation}%
Taking the back transformation $\mathbf{v}(\mathbf{r})=\mathbf{\hat{v}(%
\mathbf{r}})+\mathbf{u}$ yields the fluid velocity in the laboratory frame 
\begin{equation}
\mathbf{v}(\mathbf{r})=u\frac{a^{3}}{r^{3}}\left( \frac{1}{2}\sin \theta 
\mathbf{t}+\cos \theta \mathbf{n}\right) \ .  \label{20}
\end{equation}

Finally we determine the particle velocity $u.$ With (\ref{10}) and $\Lambda
\rightarrow 0$, the remaining condition (\ref{17}) reads 
\begin{equation*}
-v_{0}\sin \theta =\hat{v}_{\theta }|_{a+\Lambda }.
\end{equation*}%
Inserting $\hat{v}_{\theta }$ we have $u=-\frac{2}{3}v_{0}$ and, with the
expression for $v_{0}$, 
\begin{equation}
u=-\xi \frac{\varepsilon \psi _{0}^{2}}{12\eta T}\left( 1+\tau \right) A\ .
\label{18}
\end{equation}%
The maximum value of the boundary velocity $v_{B}$ occurs at $\theta =\frac{%
\pi }{2}$ and exceeds the particle velocity, i.e. the particle and the fluid
beyond the boundary layer move in opposite directions.

\subsection{Boundary layer approximation}

The expression (\ref{9}) relies on two assumptions: slow variation of the
tangential velocity in the boundary layer, $\partial _{x}^{2}\hat{v}_{x}\ll
\partial _{z}^{2}\hat{v}_{x}$, and a small shear stress beyond a distance $%
\Lambda $. Here we justify these assumptions by evaluating the quantities
from the 3D solution, and we summarize the variation of the velocity field
and the shear stress from the particle surface to distances well beyond the
boundary layer.

The parallel derivative of the velocity in the boundary layer $\partial _{x}%
\hat{v}_{x}$ matches $(1/a)\partial _{\theta }\hat{v}_{\theta }$; according
to (\ref{20}) it is of the order $v_{\theta }/a$. Comparing to the normal
derivative in the boundary layer $\partial _{z}\hat{v}_{x}\sim \hat{v}%
_{x}/\lambda $, one readily verifies $\partial _{x}\hat{v}_{x}\ll \partial
_{z}\hat{v}_{x}$, i.e., that the latter provides the dominant contribution
in terms of the small parameter $\lambda /a$; the same argument applies to
the second derivative.

In Eq. (\ref{5a}) we have used that for a flat surface, the shear stress
vanishes beyond the boundary layer. From (\ref{5b}) and (\ref{20}) one
obtains 
\begin{equation*}
\sigma _{r\theta }|_{a+\Lambda }=-\frac{2}{a}\eta v_{B},
\end{equation*}%
the shear stress is proportional to the inverse curvature radius of the
particle. The variation of the shear stress through the the boundary layer
is given by putting $z=0$ in (\ref{5a}), 
\begin{equation*}
\sigma _{xz}(0)=\frac{2}{\lambda }\eta v_{B}.
\end{equation*}%
One readily finds that $\sigma _{r\theta }|_{a+\Lambda }$ is by a factor $%
\lambda /a$ smaller than the term retained in (\ref{5a}). These relations
confirm the validity of the boundary layer approximation in the case $%
\lambda \ll a$.

\begin{figure}
\includegraphics[width=\columnwidth]{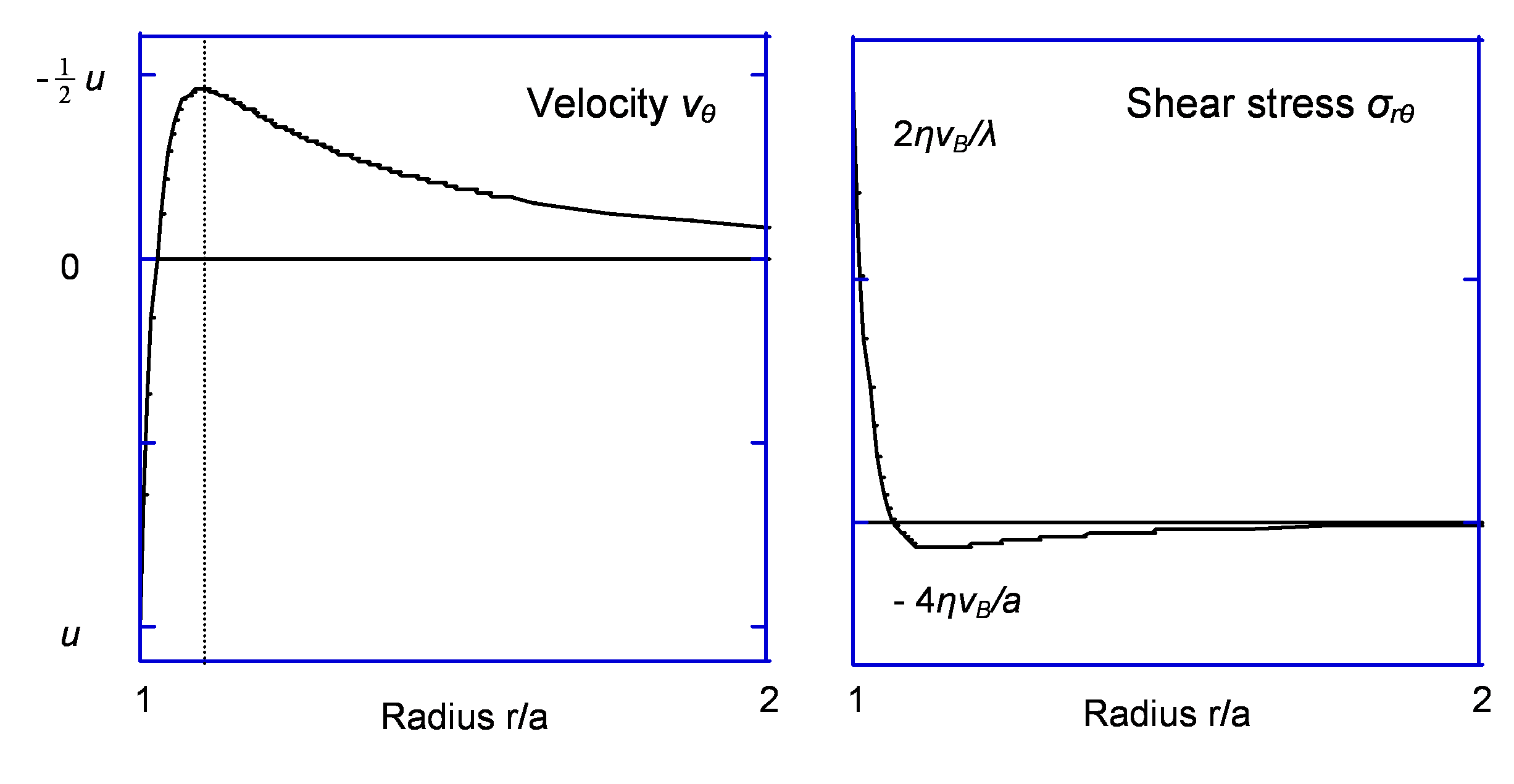}
\caption{Schematic plot
of the tangential velocity $v_{\protect\theta }$ and the shear stress $%
\protect\sigma _{r\protect\theta }$ for $\protect\theta =\frac{\protect\pi }{%
2}$ and $a\leq r\leq 2a$. The vertical dashed line indicates the thickness
of the boundary layer, i.e., the Debye length $\protect\lambda $. The
minimum and maximum values of $v_{\protect\theta }$ and $\protect\sigma _{r%
\protect\theta }$ are indicated.}
\end{figure}

In Fig.\ 2 we plot schematically the variation of both the velocity field
and the shear stress. In view of (\ref{18}) we put $u<0$, i.e., the particle
moves in the direction opposite to the thermal gradient. The left panel
shows the function $v_{\theta }(r)$ at $\theta =\frac{\pi }{2}$, i.e., in
the plane normal to the applied thermal gradient where the radial component
is zero and where the relative velocity reads $v_{B}=-\frac{3}{2}u$. At the
particle surface the fluid velocity takes the value $v_{\theta }|_{a}=u$,
increases through the boundary layer, and attains $v_{\theta }|_{a+\Lambda
}=-\frac{1}{2}u$. At larger distances, the velocity vanishes with the
characteristic power law $v_{\theta }=-\frac{1}{2}u(a/r)^{3}$. The shear
stress is shown in the right panel. Its maximum and minimum values occur at
the particle surface and beyond the boundary layer, respectively, and they
differ by a factor $\lambda /a$. At larger distances the shear stress
vanishes as $\sigma _{r\theta }^{\prime }\sim 1/r^{4}$.\ 

\section{Phoretic coefficients}

Eq. (\ref{18}) gives the phoretic velocity of the suspended particle in
terms of the applied thermal gradient and thus defines the proportionality
factor in (\ref{4a}) 
\begin{equation}
C=\xi (1+\tau )\frac{\varepsilon \psi _{0}^{2}}{12\eta T}.  \label{22}
\end{equation}%
The transport coefficient $D_{T}$ is obtained from (\ref{4b}), with the
mobility $\mu =1/(6\pi \eta a)$ and the ratio of heat capacities $\xi
=3\kappa _{S}/(2\kappa _{S}+\kappa _{P})$, 
\begin{equation}
D_{T}=\frac{k_{B}}{6\pi \eta a}+\frac{\kappa _{S}}{2\kappa _{S}+\kappa _{P}}%
\left( 1+\tau \right) \frac{\varepsilon \psi _{0}^{2}}{4\eta T}.  \label{24}
\end{equation}%
The first term depends on the particle size $a$; \ for sufficiently large
solutes it is negligible, and $D_{T}$ is independent of the particle size.
For a polymer coil, $a$ has to be replaced by the gyration radius $R$.

Most experiments study the stationnary density modulation $\delta
n/n=-(D_{T}/D)\delta T$ \ induced by the temperature inhomogeneity $\delta T$%
, and thus measure the Soret coefficient $S_{T}=D_{T}/D$ rather than the
transport coefficient $D_{T}$. With $D=k_{B}T/(6\pi \eta a)$ one has 
\begin{equation}
S_{T}=\frac{1}{T}\left( 1+\xi \left( 1+\tau \right) \frac{\pi a\varepsilon
\psi _{0}^{2}}{2k_{B}T}\right) .  \label{26}
\end{equation}%
The first term in brackets gives the ideal-gas expression $S_{T}=1/T$; the
remaining one is proportional to the particle size $a$. For solutes larger
than a few nanometers, phoretic motion due to surface forces in general
exceeds the diffusive term, i.e., the coefficient $C$ is larger than $\mu
k_{B}$. In this limit the above quantities vary with the square of the Debye
length.\ The Soret coefficient reads 
\begin{equation}
S_{T}\propto \lambda ^{2}a,  \label{28}
\end{equation}%
whereas $D_{T}\propto \lambda ^{2}$ and $u\propto \lambda ^{2}$ are
independent of the particle size.

\section{Discussion}

\subsection{Approximations}

Our results follow from a hydrodynamic treatment of the fluid surrounding a
charged particle. In view of the discrepancies with recent work discussed
below, it seems worthwhile to review the underlying assumptions.

(i) The surface charge density $\sigma $ is supposed to be constant,
resulting in a surface potential $\psi _{0}$ that depends only weakly on
temperature through the permittivity and the screening length. For most
experimental systems, the charge $\sigma $\ arises from ionic surfactants
grafted on a particle or trapped at a liquid interface.\ If the degree of
dissociation of this surfactant varied with $T$, the value of the surface
charge $\sigma $ and thus the potential $\psi _{0}$ would show an additional
temperature dependence. \ \ 

(ii) The present work relies on the validity of the Debye-H\"{u}ckel
approximation, i.e., on sufficiently small surfarce charges. Yet for several
systems the measured values of $S_{T}$ indicate effective valencies $Z$
close to the value $Z^{\ast }=(4a^{2}/\ell _{B})(1+\lambda /a)$ where the
weak-coupling assumption ceases to be valid \cite{Boc02}.

(iii) Both the hydrodynamic treatment and the electrostatics are restricted
to the leading order in powers of the parameter $\lambda /a$. For
micron-size particles this ratio is of the order of a few percent \cite%
{Duh06,Duh06a} yet approaches unity for micelles and water-in-oil droplets
of a few nanometers \cite{Pia02,Vig07,Put07}.

(iv) In Eq. (\ref{12}) we have supposed that the charge distribution in the
double layer is not affected by the thermal gradient, i.e., we have
neglected polarization effects; preliminary work \cite{Dho08} indicates that
polarization corrections are of the order of $\lambda /a$ and thus may
safely be neglected for large particles..

\subsection{Hydrodynamic boundary conditions}

Comparison with the discussion of the charged double layer at the end of 
\cite{Wue07} reveals that Eq. (\ref{26}) differs by a factor $\lambda /a$.
This discrepancy arises from the boundary conditions for the velocity field
at the solid-fluid interface. The present work is based on the no-slip
boundary conditions (\ref{1c}), i.e., both tangential and normal components
of the velocity are continuous, and in particular $\hat{v}_{x}(z=0)=0$ in (%
\ref{5c}). On the contrary, Ref. \cite{Wue07} uses perfect-slip conditions,
corresponding to the reference value $\sigma _{xz}(0)=0$ of the surface
stress, 
\begin{equation}
\sigma _{xz}(z)=-\int_{0}^{z}dz^{\prime }f_{x},
\end{equation}%
instead of (\ref{5a}). Inserting the force field (\ref{7}) one readily finds
the stress on the fluid beyond the boundary layer, $\sigma _{xz}(\Lambda
)=-(\sigma ^{2}/4\varepsilon \lambda T)(1+\tau )\partial _{x}T$, which
confirms Eq. (15) of \cite{Wue07}. The resulting expression for the Soret
coefficient exceeds the present one\ by a factor $a/\lambda $. On the other
hand, when evaluating the surface stress from Eq. (\ref{5b}) with the
no-slip boundary velocity $v_{B}$, we find $\mathbf{n}\cdot \mathbf{\sigma }%
\cdot \mathbf{t}=-2\eta v_{B}/a$; inserting this in Eq. (6) of \cite{Wue07},
one recovers the above results (\ref{22}-\ref{26}).

Thus the conditions of zero tangential velocity $(\hat{v}_{x}(0)=0)$ and
zero shear stress $(\sigma _{xz}(0)=0)$ result in Soret coefficients that
differ by a factor $\lambda /a$. This means that a much stronger Soret
effect is expected for suspensions that satisfy slip boundary conditions,
thus illustrating the importance of the properties of the particle-solvent
interface. We note that the data on AOT/water/oil microemulsions, SDS
micelles, and polystyrene nanoparticles \cite{Pia02,Vig07,Put07} rather
agree with the present result $S_{T}\propto a$ based on no-slip conditions,
whereas those on micron size polystyrene beads would match the law $%
S_{T}\propto a^{2}$ that follows from slip \cite{Duh06,Duh06a}.

Available data suggest that significant slip may occur at hydrophobic
interfaces \cite{Lau07}, with slip lengths ranging from a few nanometers to
a micron. Perfect slip as assumed in \cite{Wue07} occurs if the particle
size is smaller than the slip length.

\subsection{Previous theoretical work}

Following Smoluchowski's argument for electrophoresis \cite{Smo18},
Ruckenstein suggested a size-independent phoretic velocity $u$ \cite{Ruc81},
implying a Soret coefficient $S_{T}\propto \lambda ^{2}a$, which was
confirmed more recently by Refs. \cite{Mor99,Par04} and agrees with our Eq. (%
\ref{28}). Regarding the prefactors, Refs. \cite{Ruc81,Par04} discuss only
the dominant behavior and do not account for the modified temperature
gradient (\ref{11}).\ Our result confirms that of Morozov \cite{Mor99} in
the limit $\lambda \rightarrow 0$ and for weak coupling; we can make no
statement concerning the negative Soret coefficient derived in \cite{Mor99}
for strong charges.

More recent work took the thermal force as the gradient of the charging
energy of the double layer \cite{Duh06,Duh06a,Bri03,Fay05,Dho07}. This
assumption results in dependencies of the Soret coefficient, $S_{T}\propto
\lambda a^{2}$, that significantly differ from those given above. In order
to point out the main differences, we rewrite Eq. (\ref{26}) in terms of the
charging energy $E_{C}=\frac{1}{2}Q\psi _{0}$; with the relation $\psi
_{0}=(Q/4\pi \varepsilon )(\lambda /a^{2})$ one has%
\begin{equation}
S_{T}=\frac{1}{T}\left( 1+\frac{\xi }{8}\left( 1+\tau \right) \frac{\lambda 
}{a}\frac{Q\psi _{0}}{k_{B}T}\right) .
\end{equation}%
Comparison with, e.g., Eq. (44) of \cite{Dho07} in the limit of thin
boundary layers, reveals that our expression is by a factor $\lambda /a$
smaller than that obtained from the gradient of the charging energy.

The ratio $\xi $ of thermal conductivities of solute and solvent is missing
in most previous works. This factor $\xi $\ accounts for the local
distortion of the temperature field $T(\mathbf{r})$; e.g., for example, if
the particle is a good heat conductor, the temperature in its vicinity is
almost constant, and its gradient is small. Depending on the thermal
properties of solute and solvent, the factor $\xi $ may considerably reduce
the Soret effect. An full discussion of suspensions of metal particles is
given in \cite{Gid95}. \ 

Finally we note that the present approach differs from Derjaguin's model 
\cite{Der87} which is based on enthalpy transport in a thermal gradient.
This is most obvious when comparing the boundary velocity in Eq. (\ref{9})
to the expression given in Chapts. 7 and 11 of \cite{Der87} or in the review
by Anderson \cite{And89}.

\subsection{Experiments}

Available experimental findings \cite{Pia02,Duh06,Duh06a,Vig07,Put07}
diverge with respect to the dependencies of the Soret coefficient on Debye
length $\lambda $ and particle size $a$. At present it is not clear whether
the measured Soret effect varies linearly or with the square of the Debye
length; see e.g. the discussion in \cite{Pia02,Dho07}. When comparing these
measurements with the present or previous theoretical results, one should
keep in mind that discrepancies could arise from the weak-charge assumption;
is it by no means clear that the charged colloidal systems discussed above
satisfy the condition of weak coupling.

Regarding the dependence on the particle size $a$, very recent studies on
AOT/water/oil microemulsions \cite{Vig07} and carboxyl functionalized
polystyrene particles \cite{Put07} show a linear dependence on the particle
size, $S_{T}\propto a$, in the range of a few nanometers up to several tens
of nm. Thus these experiments would agree with \cite{Ruc81,Par04} and our
Eq. (\ref{28}) which is based on hydrodynamics with no-slip boundary
conditions. On the other hand, a quadratic power law $S_{T}\propto a^{2}$
has been reported for micron-size polystyrene particles \cite{Duh06,Duh06a}.
Such a behavior has been obtained theoretically from the model based on the
gradient of the charging energy \cite{Duh06,Duh06a,Bri03,Fay05,Dho07}, and
from the boundary layer approach with perfect slip conditions \cite{Wue07}.

A.W. acknowledges stimulating and helpful discussions with R.\ Piazza, D.\
Braun, J.\ Dhont, and S.\ Wiegand.

\end{document}